\documentclass[draft]{iopart}

\usepackage[final]{graphicx}
\usepackage{cite}

\begin{document}

\title[Influence of equation of state...]{Influence of equation of state on interpretation \\ of electrical conductivity measurements in strongly coupled tungsten plasma}

\author{S I Tkachenko, P R Levashov, and K V Khishchenko}

\address{Institute for High Energy Densities, Joint Institute for High Temperatures, Russian Academy of Sciences, Izhorskaya 13/19, Moscow 125412, Russia}
\ead{svt@ihed.ras.ru}
\begin{abstract}
We study the influence of equation-of-state (EOS) model on the
interpretation of electrical conductivity measurements in strongly coupled
plasma of tungsten by Korobenko {\em et al.\/}\ (2002 {\em Plasma Physics Reports\/} {\bf 28}(12) 1008--1016).
Three different semiempirical EOS models for tungsten are used.
Discrepancies in obtained thermodynamic parameters and specific resistivity
values as compared with calculation results of Korobenko {\em et al.\/}\ are analysed.
\end{abstract}

\pacs{64.30.+t, 72.15.Cz, 52.25.$-$b}
%\maketitle

\section{Introduction}

Electrical explosion of wires or foils is an effective way to
study thermophysical properties of matter in a wide range of
densities and temperatures
\cite{Lebedev:Savvat:1984,Gathers:1986}. This is one of a few
methods, which allows one to obtain both thermodynamic properties and kinetic
coefficients in the same experiment. For example, electrical
resistance can be calculated from experimental time dependencies of the heating
current and voltage. To determine specific properties it should be
known also the cross-section area of the conductor as a function of
time. As geometric sizes of the sample may be not measured in
the process of expansion it is reasonable to use the results of
numerical simulation. In this case calculated properties of matter
are determined, in particular, by an equation-of-state (EOS) model. In the present work we study the influence of the 
EOS model to the values of electrical conductivity of strongly coupled tungsten
plasma based on data from Korobenko {\em et al.\/}\ \cite{Korobenko:Rakhel:2002}.

\section{Description of experiment}

The experiments on electrical conductivity measurements \cite{Korobenko:Rakhel:2002} were
carried out in a plane geometry. 
A tungsten foil stripe with the length $l_z = 10$~mm, width $h = 1.5$~mm 
and thickness $2a =20$~$\mu$m was placed between two glass
plates with the thickness $a_1 = 5$~mm. Side slits were shielded
with thin mica stripes. In the experiment under consideration the skin layer thickness
$\delta$ is significantly larger than the foil thickness.
Cartesian coordinate system is introduced as follows: $x$-axis is
perpendicular to the foil plate, $y$-axis is directed along the
smaller side of the foil, and $z$-axis --- along the bigger side.
In 1D process the foil expands along the $x$-axis, the magnetic
induction $B$ is directed along the $y$-axis, and the heating
current $I$ as well as the electric field intensity $E$ are directed
along the $z$-axis.

The foil was heated by the impulse of current; the time
dependencies of the current through the sample $I(t)$ and voltage
drop $U(t)$ were registered (\fref{fig_i-t_u-t}). Then it was
calculated the resistive part of the voltage drop $U_R(t)$,
electrical resistance $R(t) = U_R(t)I^{-1}(t)$ and Joule heat $q(t)$.
Other values required for conductivity calculation can be obtained by means of numerical simulation.
Assuming that the current density $j$ is distributed uniformly
over the cross-section of the foil and depends only on time, i.e.
$j(t) = I(t)S^{-1}(t)$, where $S(t) = 2a(t)h$, 
from the Maxwell equation $j(t) = \mu^{-1}\partial B/\partial x$ (SI system of
units is used, $\mu$ is the magnetic permeability) one can calculate $B(t,x) = \mu I(t)xS^{-1}(t)$.
So it is possible to determine the $x$--$t$-dependencies of foil parameters
as a numerical solution of only a set of hydrodynamic
equations with the Ampere force $jB = \mu I^2(t)xS^{-2}(t)$ and energy input 
$jE=U(t)I(t)V^{-1}(t)$, where $V(t) = S(t)l_z$ is the foil volume.

The results of calculation by
such a technique not allowing for magnetic field diffusion were presented in work \cite{Korobenko:Rakhel:2002}.

\section{Modeling}

Assuming that spatial perturbations of the sample form are small
and electron and ion temperatures are equal each other, the set
of 1D magnetohydrodynamic (MHD) equations in Lagrangian description for the foil heating can be represented as follows:

\begin{equation}
  dm/dt = 0,
  \label{mass}
\end{equation}
\begin{equation}
  \rho dv/dt = -\partial P/\partial x - (2\mu)^{-1}\partial B^2/\partial x,
  \label{momentum}
\end{equation}
\begin{equation}
  \rho d\varepsilon /dt = -P \partial v/\partial x +
  \partial ( \kappa \partial T/\partial x)/\partial x + j^2/\sigma,
  \label{energy}
\end{equation}
\begin{equation}
  d(\mu B)/dt = \partial (\sigma^{-1} \partial B/\partial x)/\partial x,
  \label{magnetic}
\end{equation}
where $m$ is the mass, $v$ is the particle velocity, $\rho$ is the density, 
$T$ is the temperature,
$P$ is the pressure,
$\varepsilon$ is the specific internal energy,
$\sigma$ is the electrical conductivity,
$\kappa$ is the thermal conductivity.
Initial conditions for the set of equations (\ref{mass})--(\ref{magnetic}) are
written as follows: $\rho(x, 0) = \rho_0$, $v(x, 0) = 0$, $T(x, 0)
= T_0$, $B(x, 0) = 0$.
The conditions
on the symmetry plane $x = 0$ and
on the surface $x = a(t)$ of the foil,
as well as on the outer boundary of the glass plate $x = a_1$ are as follows:
$v(0,t) = 0$, $v(a,t) = da/dt$, $v(a_1,t) = 0$,
$B(0,t) = 0$, $B(a,t) = \mu I(t)/2h$,
$\partial T/\partial x|_{x = 0} = 0$, 
$\partial T/\partial x|_{x = a - 0} = \partial T/\partial x|_{x = a + 0}$,
$T(a_1,t) = T_0$,
$\partial P/\partial x|_{x=0} = 0$,
$P(a-0,t) = P(a+0,t)$,
$P(a_1, t) = P_0$.
Here $\rho_0$, $T_0$, and $P_0$ correspond to normal conditions.

We used three different EOS models for tungsten \cite{Young:1977:Report,Khishchenko:Elbrus2005:EOS,Khishchenko:TPL2005:EOS}.
Semiempirical multi-phase EOS \cite{Khishchenko:Elbrus2005:EOS} in a form of functions 
$P=P(\rho,T)$ and $\varepsilon=\varepsilon(\rho,T)$ (EOS1) takes into account the effects of
high-temperature melting, evaporation, and ionization. 
This EOS agrees with the collection of experimental data on static and
shock compression as well as on adiabatic and isobaric expansion
of the metal, see details in \cite{Khishchenko:Elbrus2005:EOS}. 
Caloric EOS \cite{Khishchenko:TPL2005:EOS} in a functional form $P=P(\rho,\varepsilon)$ (EOS2) 
neglects phase transitions; however it describes available shock-wave experiments within a good accuracy. 
The soft-sphere EOS \cite{Young:1977:Report} as functions 
$P=P(\rho,T)$ and $\varepsilon=\varepsilon(\rho,T)$ with coefficients from \cite{Hess:IJT:1999}
(EOS3) considers evaporation of metal and 
has been calibrated using isobaric expansion experiments 
but does not take into account melting and gives understated density at normal temperature and pressure. 

%CBP
As EOS1 allows for more effects and agrees with wider collection of data 
including the region of parameters of the considered experiment \cite{Korobenko:Rakhel:2002} 
this model is assumed to be more reliable than EOS2 and EOS3. 
However that may be, the correct EOS 
can be chosen (not necessarily amongst three used models) 
only in the case of direct thermodynamic measurements in the range of interest.
%EOC

To describe the properties of glass we used caloric EOS $P=P(\rho,\varepsilon)$  \cite{Khishchenko_etal_SCCM1995}.

The conductivity of tungsten was determined by the relation 

\begin{equation}
 \sigma = I(t)l_z U^{-1}(t)S^{-1}(t)
  \label{sigma}
\end{equation}

\noindent using the experimental dependencies $I(t)$ and $U(t)$
\cite{Korobenko:Rakhel:2002} except for the stage of heating up to $T=10$~kK. In case of EOS1 at low temperatures we used the
semiempirical formulas \cite{Knoepfel:1970,Tkachenko:TVT:2001,Tkachenko:IJT:2005} 
for the electrical conductivity $\sigma=\sigma(\rho, T)$ taking into account melting effect
instead of experimental functions because of noise on the measured time dependence of voltage at the initial stage. 
The thermal conductivity in case of EOS1 was calculated according to the Wiedemann--Franz law,
$\kappa=k_{WF}T\sigma$, where $k_{WF}$ is the Wiedemann--Franz constant.
In cases of EOS2 and EOS3 during the initial stage 
we used time dependence of voltage $U(t)$ obtained in numerical modeling with EOS1. 
Thus in these cases the electrical conductivity was determined according to \eref{sigma}
during the whole heating process. The thermal conductivity effects in cases of EOS2 and EOS3 were neglected. 

\section{Results}

\begin{figure}
\centering
  \includegraphics[width=0.9\textwidth]{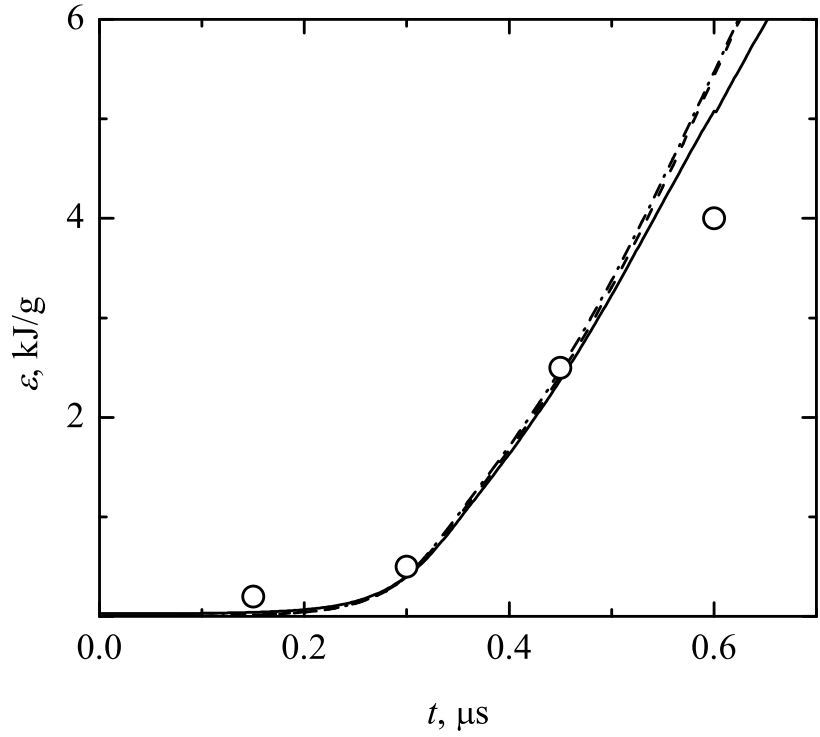}
  \caption{
    Specific internal energy versus time in the foil during heating 
calculated basing on measurements of Korobenko {\em et al.\/}\ \cite{Korobenko:Rakhel:2002}:
circles are from simulations of Korobenko {\em et al.\/}\ \cite{Korobenko:Rakhel:2002}, 
lines correspond to results of present work simulations 
in case of EOS1 (solid line), EOS2 (dashed line), and EOS3 (dash-dotted line).
  }
  \label{osc_ener}
\end{figure}

We carried out a number of simulations of the experiment using 1D
MHD model as described in the previous section. 
In \fref{osc_ener} shown are time dependencies of the specific internal energy $\varepsilon(t)$ resulting from numerical modeling 
with three above-mentioned EOS. One  can see that all three curves $\varepsilon(t)$ are very close during the initial heating stage.

\begin{figure}
\centering
  \includegraphics[width=0.9\textwidth]{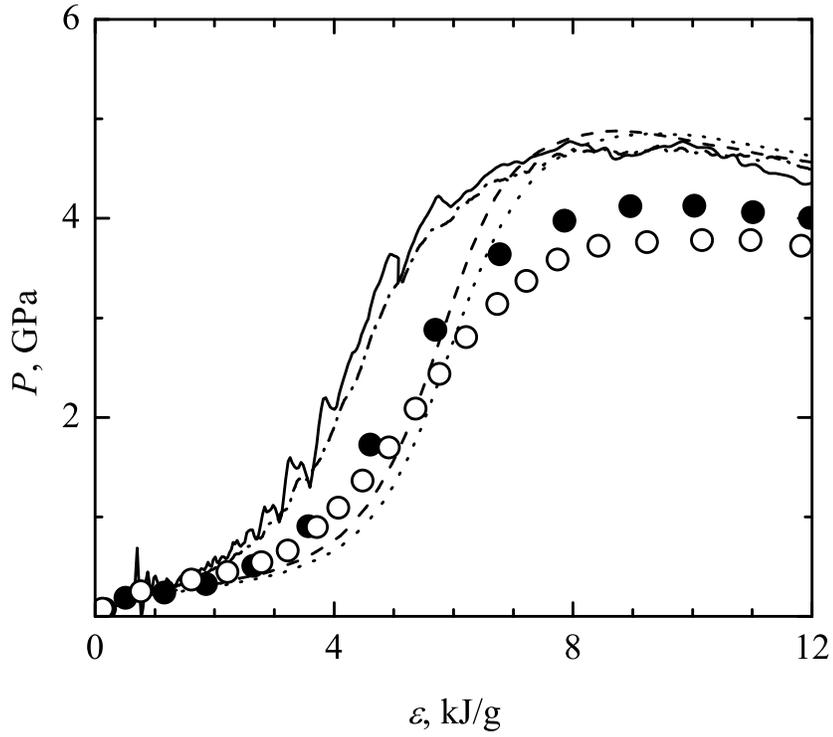}
  \caption{
    Pressure versus specific internal energy in the foil during heating
from calculations based on measurements of Korobenko {\em et al.\/}\ \cite{Korobenko:Rakhel:2002}:
open circles are from simulations of Korobenko {\em et al.\/}\ \cite{Korobenko:Rakhel:2002}, 
lines and solid circles denote results of numerical simulations of present work in case of EOS1 (solid and dash-dotted lines, for layers $x=0$ and $a(t)$ correspondingly), 
EOS2 (dashed and dotted lines, $x=0$ and $a(t)$ correspondingly), and EOS3 (solid circles, $x=0$).
}
  \label{osc_pres}
\end{figure}

The calculated pressure at the symmetry
plane and at the foil surface depending on the specific
internal energy at the same layers is shown in \fref{osc_pres} in comparison with results from simulations of Korobenko {\em et al.\/}\ \cite{Korobenko:Rakhel:2002}.
It can
be seen that the melting process leads to oscillations of pressure $P(\varepsilon)$
near the symmetry plane of the foil (see curve at $x=0$ for EOS1 in \fref{osc_pres}). 
If melting is neglected (like in calculations with EOS2 and EOS3 models) 
the pressure dependencies $P(\varepsilon)$ are smooth. The dynamics of the process is to some extent determined by the EOS model: 
after the melting EOS1 gives the fastest pressure rise during expansion, 
EOS2~--- the slowest one. 
At the later stage of the heating EOS1 and EOS2 result in close pressure values while
EOS3 shows 15\% lower pressures. 

The EOS model used in work \cite{Korobenko:Rakhel:2002} for the interpretation of
experimental data is based upon the soft-sphere EOS
\cite{Young:1977:Report} and takes into account ionization effects
according to the mean atom model \cite{Basko:1985}. 
This EOS is unpublished and this fact complicates the qualitative analysis of distinctions. 
Nevertheless, \fref{osc_pres} shows that the
calculated pressure \cite{Korobenko:Rakhel:2002} in the process of
foil heating is always lower than in present work; the same
situation is observed for temperature. 
The 
simulation with EOS3 gives the closest result to that of the
work \cite{Korobenko:Rakhel:2002}; this coincidence can be explained by
similar EOS models used in these calculations. 

\begin{figure}
\centering
  \includegraphics[width=0.9\textwidth]{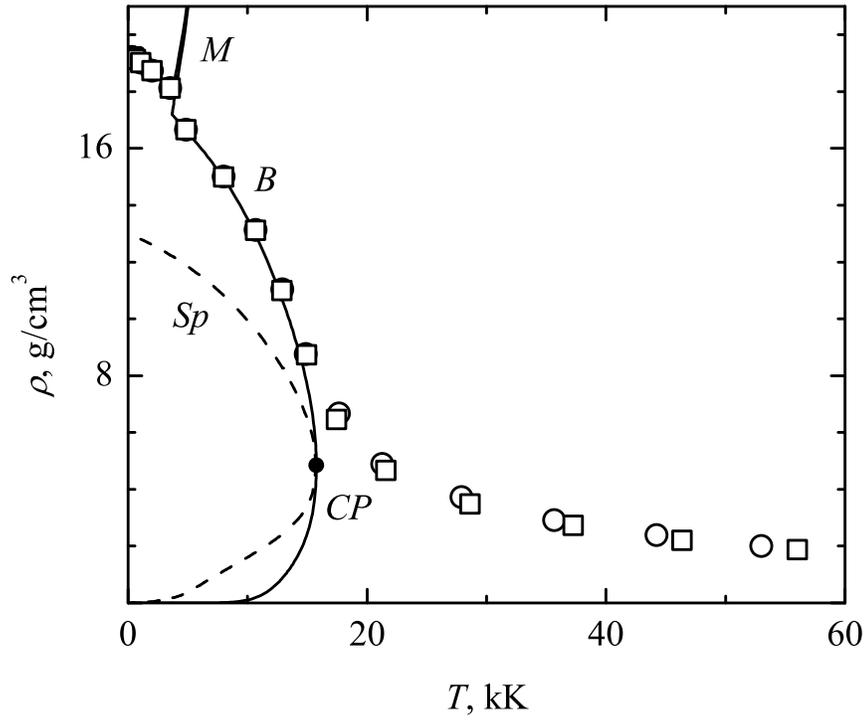}
  \caption{Phase diagram of tungsten \cite{Khishchenko:Elbrus2005:EOS} and phase trajectories:
    \textit{M} is the melting region, \textit{B} is the boundary of liquid--gas transition region, \textit{CP} is the critical point,
    \textit{Sp} are spinodals of the liquid and gas phases, circles denote states at the symmetry plane of the foil
    during heating, each square corresponds to state on the foil surface at the same moment as the nearest circle.
  }
  \label{phase_diagram}
\end{figure}

\begin{figure}
\centering
  \includegraphics[width=0.9\textwidth]{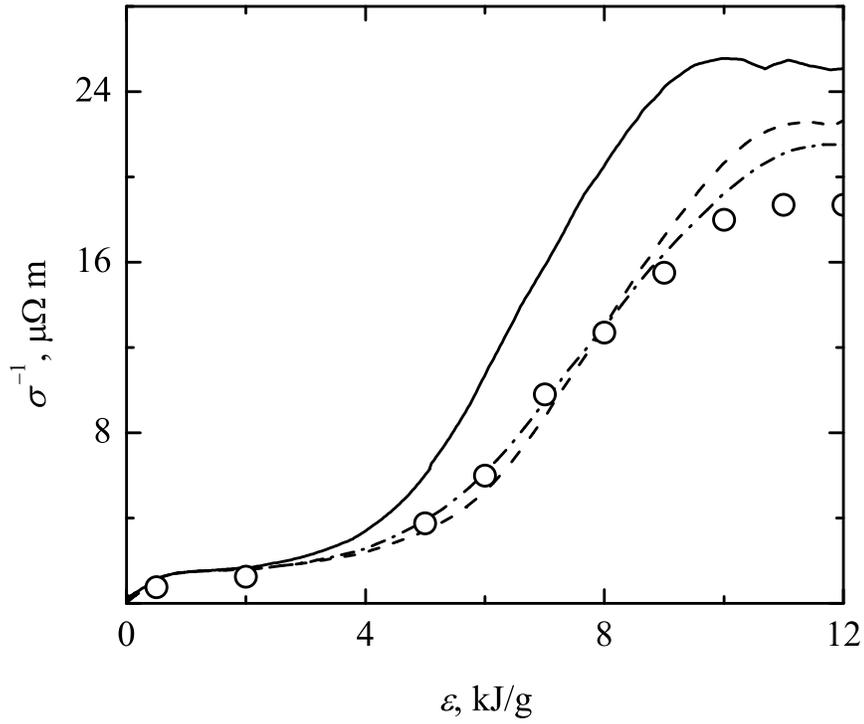}
  \caption{
    Specific electrical resistivity of tungsten versus specific internal energy in the foil during heating
from calculations based on measurements of Korobenko {\em et al.\/}\ \cite{Korobenko:Rakhel:2002}:
circles are from simulations of Korobenko {\em et al.\/}\ \cite{Korobenko:Rakhel:2002}, 
lines correspond to results of present work simulations 
in case of EOS1 (solid line), EOS2 (dashed line), and EOS3 (dash-dotted line).
  }
  \label{sigma_e}
\end{figure}

One can see in \fref{osc_pres} that parameters in the foil are
distributed homogeneously except for the moment
of melting which is clearly distinguishable by pressure
oscillations. 
%CBP
After melting thermodynamic states of the foil though sometimes very close to the binodal are always in liquid or supercritical plasma state (\fref{phase_diagram}). 
%EOC
However, inhomogeneity in temperature distribution
appears at the late stage of the expansion process. For example acording to modeling with EOS1 the scale of
this inhomogeneity can be distinctly seen in \fref{phase_diagram}
where the thermodynamic tracks of different layers of the foil are
shown. 
%CBP
%From the same figure one can see that the foil is always at
%higher temperature than that of the liquid--gas transition at the same
%density. In other words, the foil is always in liquid or
%supercritical plasma state after melting. 
%EOC
Distinctions in the methodology
of simulation and description of thermodynamic properties of
tungsten lead to systematically higher values of electrical resistivity
in our interpretation (maximum excess is about 60\% for simulation with EOS1) 
than after Korobenko {\em et al.\/}\ \cite{Korobenko:Rakhel:2002} (\fref{sigma_e}).

\begin{figure}[h]
\centering
  \includegraphics[width=0.9\textwidth]{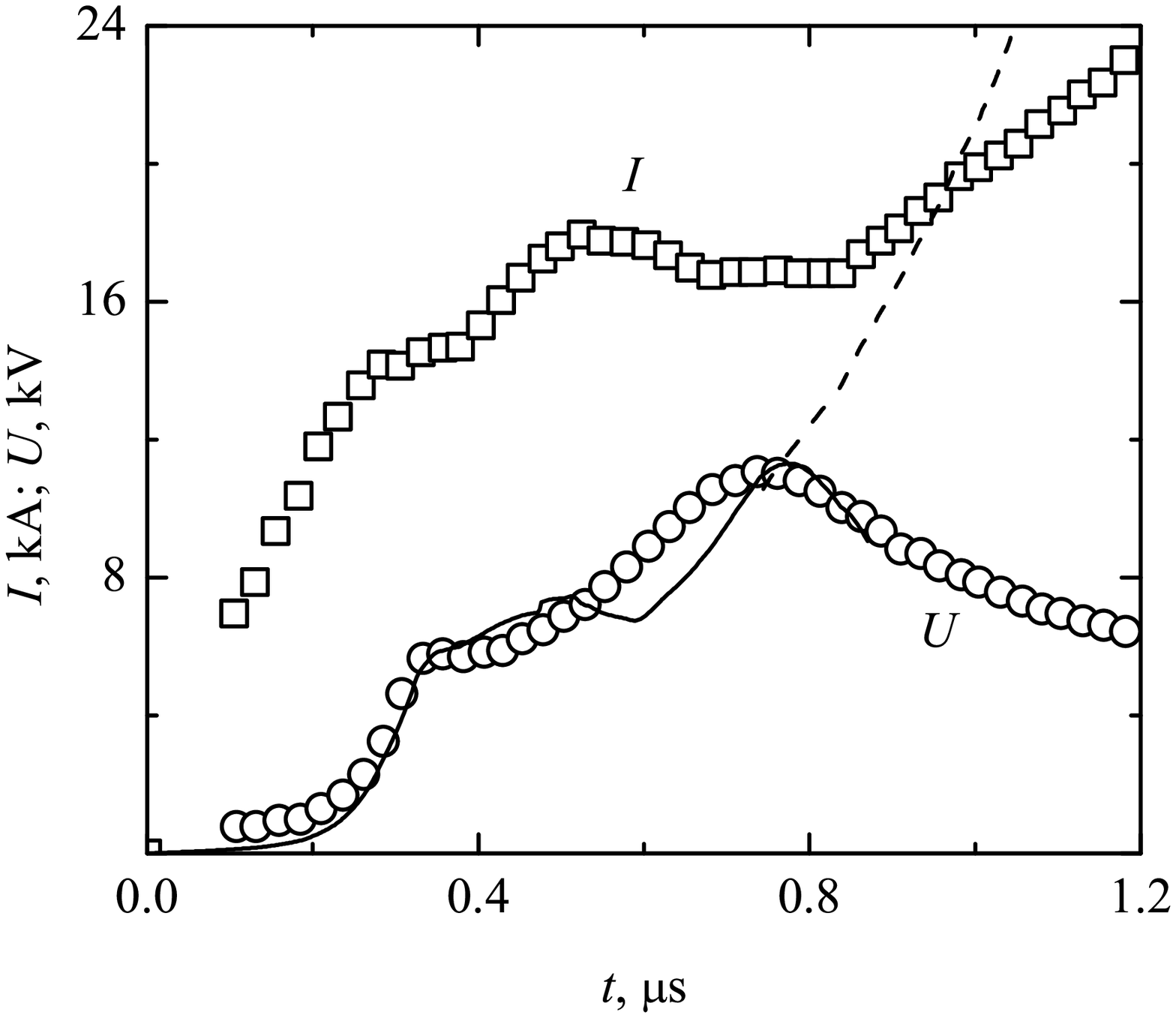}
  \caption{Current and voltage versus time: $I$ is the current, $U$ is the voltage, 
    markers correspond to data from measurements of Korobenko {\em et al.\/}\ \cite{Korobenko:Rakhel:2002},
    lines denote results of numerical simulation of present work with conductivity model \cite{Oreshkin:TP:2004}
    taking into account breakdown effect (solid line) as well as disregarding breakdown (dashed line).
  }
  \label{fig_i-t_u-t}
\end{figure}

It is worth to mention that during the heating process the measured voltage 
%CBP
%drops sharply 
begins to drop
%EOC
at time $t \sim 750$~ns and soon (at $t\sim 850$~ns) current begins to rise (see \fref{fig_i-t_u-t}).
This effect in work \cite{Korobenko:Rakhel:2002} is connected with the
beginning of transition into ``dielectric'' (plasma) state.
%CBP
Actually there is a drop in resistance at the later stage of the experiment \cite{Korobenko:Rakhel:2002}, 
however there is no noticeable change in specific resistance (see \fref{sigma_e}, $\varepsilon>10$~kJ/g).
From \fref{phase_diagram} it can be seen that supercritical (plasma) state of tungsten, $\rho \leq \rho_{cr}$, 
is reached at temperature $T\sim 20$~kK, which corresponds to time $t\sim 0.6$~$\mu$s and specific internal energy $\varepsilon\sim 5$~kJ/g. 
It is known that the character of electrical conductivity changes from metal-like to plasma-like near the critical
density \cite{Oreshkin:TP:2004}, $\rho_{cr}=4.854$~g/cm$^3$ for tungsten according to EOS1, 
so one can expect that the resistivity dependence should change its behavior at values of internal energy $\varepsilon \sim 5$~kJ/g. 
As this is not the case in \fref{sigma_e}
we offer an alternative explanation of the ``saturation'' of electrical conductivity. Namely we 
assume that a breakdown of interelectrode gap takes place along the glass surface at $t \sim 750$~ns 
(the density and specific internal energy of tungsten foil at this moment are $\rho \sim 2$~g/cm$^{3}$ and $\varepsilon \sim 8$~kJ/g).

%EOC
We tried to
reproduce the experimental time dependence of voltage using
time dependence of current as input data together with wide-range conductivity
models for tungsten \cite{Oreshkin:TP:2004,Apfelbaum:HT:2003} and EOS1 model. The
results of simulation with the conductivity model
\cite{Oreshkin:TP:2004} are shown in \fref{fig_i-t_u-t}. We could
well describe the voltage data up to $t \sim 500$~ns
and satisfactory up to $t \sim 750$~ns, but when the voltage began to drop
in the experiment it continued to rise in our calculations. It is
possible to reproduce the maximum around $t \sim 750$~ns on the experimental voltage
dependence only by use of a breakdown model in the
simulation. 
%CBP
%Namely from \fref{fig_i-t_u-t} we can estimate the interelectrode gap breakdown development time (from 
%voltage maximum to the beginning of current rise) as 100~ns and the rate as 100~km/s. These
%rates are typical at a development of streamer discharges in dense media \cite{Lesaint:1998}. 
Assuming
%EOC
that the breakdown occurs on the
boundary between the glass plate and tungsten foil we increased
the electrical conductivity value
%$\sigma$ 
in region $0.95 \le x/a(t) \le 1$ 
%$0.95a(t) \le x \le a(t)$ 
by an order of magnitude linearly during time interval from $t=750$ to 900~ns. 
It can be easily seen in \fref{fig_i-t_u-t} that in this case we have much
better agreement with experimental time dependence of voltage.
Simulations with electrical conductivity model
\cite{Apfelbaum:HT:2003} showed much more worse results than that
with the model \cite{Oreshkin:TP:2004} and are not displayed in
\fref{fig_i-t_u-t}. 

Thus either a breakdown occurs during the expansion of the foil or one can 
formulate an electrical conductivity model, which will be able to reproduce the voltage maximum. 
To solve this dilemma is an aim of a future work.

\section{Conclusions}

In this work we have analyzed the experiment on electrical conductivity measurements of strongly coupled tungsten plasma under heating by the current pulse. 
We have used 1D MHD simulation and different EOS models to study distribution of parameters in the foil. 
We have also tried to reproduce experimental voltage time dependence using two electrical conductivity models. 

We can conclude that pressure, density and temperature are distributed almost homogeneously across the foil except for the melting stage of the process. 
The dynamics of the heating and expansion is determined by the EOS model giving rise to distinctions in electrical resistivity values up to 60\%. 

Moreover, the last stage of the experiment is probably influenced by the shunting breakdown of the interelectrode gap. 
These facts indicate that even in the case of the foil heating regime \cite{Korobenko:Rakhel:2002}, 
where certain efforts have been taken for achievement of homogeneous distribution of thermophysical parameters to simplify interpretation, 
there are still open problems to treat experimental data. 
We believe that further investigations of thermodynamic and transport properties of tungsten plasma will be helpful 
for the creation of adequate wide-range EOS and electrical conductivity models.

\section*{Acknowledgements}
The authors are grateful to V.\,N.~Korobenko, A.\,D.~Rakhel, and
A.\,I.~Savvatimskiy for valuable comments. Thanks are also owed to
V.\,I.~Oreshkin and E.\,M.\,Apfelbaum for providing with tungsten electrical
conductivity tables. 

This work was done under financial
support of the Russian Foundation for Basic Research, Grants No.\,04-02-17292, 05-02-16845, and 05-02-17533.

\clearpage

\section*{References}

\bibliographystyle{iopart-num}   % if natbib is available

\bibliography{refs}

%  \typeout{******************************************}
%  \typeout{** Please run "bibtex \jobname" to optain}
%  \typeout{** the bibliography and then re-run LaTeX}
%  \typeout{** twice to fix the references!}
%  \typeout{******************************************}
\hyphenation{Post-Script Sprin-ger}
\providecommand{\newblock}{}

\clearpage

\section*{List of figures}

\end{document}